\def\Aw8   {$A_{\omega}^{\delta=0.8}$\ }
\def\wt {$\omega(\theta)$}

\def\rmag {{$r^*$~}}
\def\etc {{\it etc.}}
\def\eg {{\it e.g.}\ }

\def\etal{{\it et al.}}

\documentclass[preprint2]{aastex}

\slugcomment{ApJ, in press}

\lefthead{Infante et al.}
\righthead{Clustering of Galaxy Pairs}

\begin{document}

\title{The Angular Clustering of Galaxy Pairs}

\author{Leopoldo Infante\altaffilmark{1,2}, 
Michael A. Strauss\altaffilmark{2}, 
Neta A. Bahcall\altaffilmark{2}, 
Gillian R. Knapp\altaffilmark{2}, 
Robert H. Lupton\altaffilmark{2}, 
Rita S.J. Kim\altaffilmark{2,3},
Michael S. Vogeley\altaffilmark{2,4},
J. Brinkmann\altaffilmark{5}, 
Istvan Csabai\altaffilmark{6,7}, 
Masataka Fukugita\altaffilmark{8}, 
Gregory Hennessy\altaffilmark{9},
\v{Z}eljko Ivezi\'c\altaffilmark{2},
Don Q. Lamb\altaffilmark{10},
Brian C. Lee\altaffilmark{11},
Jeffrey R. Pier\altaffilmark{12},
D.G. York\altaffilmark{10,13}
}

\altaffiltext{1}{Departamento de Astronom\'{\i}a y Astrof\'{\i}sica,
P. Universidad Cat\'olica de Chile, Casilla 306, Santiago 22, Chile}
\altaffiltext{2}{Princeton University Observatory, Princeton, NJ
08544-0001, USA}
\altaffiltext{3}{Department of Physics and Astronomy, The Johns Hopkins University,
   3701 San Martin Drive, Baltimore, MD 21218, USA}
\altaffiltext{4}{Department of Physics, Drexel University,
  3141 Chestnut St., Philadelphia, PA 19104}
\altaffiltext{5}{Apache Point Observatory, P.O. Box 59,
Sunspot, NM 88349-0059}
\altaffiltext{6}{
Department of Physics and Astronomy, The Johns Hopkins University,
   3701 San Martin Drive, Baltimore, MD 21218, USA}
\altaffiltext{7}{Department of Physics of Complex Systems,
E\"otv\"os University,
   P\'azm\'any P\'eter s\'et\'any 1/A, Budapest, H-1117, Hungary}
\altaffiltext{8}{Institute for Cosmic Ray Research, University of
Tokyo, Midori, Tanashi, Tokyo 188-8502, Japan}
\altaffiltext{9}{U.S. Naval Observatory, 
3450 Massachusetts Ave., NW, 
Washington, DC  20392-5420}
\altaffiltext{10}{The University of Chicago, Astronomy \& Astrophysics
Center, 5640 S. Ellis Ave., Chicago, IL 60637}
\altaffiltext{11}{Fermi National Accelerator Laboratory, P.O. Box 500,
Batavia, IL 60510}
\altaffiltext{12}{U.S. Naval Observatory, Flagstaff Station, 
P.O. Box 1149, Flagstaff, AZ  86002-1149}
\altaffiltext{13}{Enrico Fermi Institute, 5640 S. Ellis Ave., Chicago, IL 60637}

\begin{abstract}

We identify close pairs of galaxies from $278\,\rm deg^2$ of Sloan
Digital Sky Survey commissioning imaging data.  The pairs are drawn
from a sample of 330,041 galaxies with $18 < r^* < 20$.  We determine
the angular correlation function of galaxy pairs, and find it to be
stronger than the correlation function of single galaxies by a factor
of $2.9\pm0.4$.  The two correlation functions have the same
logarithmic slope of 0.77.  We invert Limber's equation to estimate
the three-dimensional correlation functions; we find clustering
lengths of $r_0= 4.2\pm0.4\,h^{-1}$ Mpc for galaxies and $7.8\pm0.7~h^{-1}$
Mpc for galaxy pairs.  
These results agree well with the global richness
dependence of the correlation functions of galaxy systems.

\end{abstract}


\section{Introduction}
\label{sec:intro}

Clusters are observed to be more
strongly clustered than galaxies (Bahcall \& Soneira 1983; Bahcall
1988; Postman \etal\ 1992; Croft \etal\ 1997;
Abadi \etal\ 1998). Moreover, the
clustering strength of groups and clusters has been shown to increase
with  richness
(Bahcall \& Soneira 1983, Szalay \& Schramm 1985, Bahcall \& West 1992). Groups of galaxies have a smaller
correlation length than that of rich clusters (Merch\'an \etal\ 2000) but
larger than that of individual galaxies (Connolly \etal\ 1998, Loveday
\etal\ 1996, Infante \& Pritchet 1995).  A number of authors
have examined the correlation function of groups of galaxies from
relatively shallow redshift surveys (Jing \& Zhang 1988; Maia \& da
Costa 1990; Ramella \etal\ 1990; Trasarti-Battistoni \etal\ 1997;
Girardi, Boschin \& da Costa 2000; Merch\'an, Maia, \& Lambas 2000),
finding that the groups exhibit a correlation function somewhat stronger
than that of galaxies.  The richness dependence of
the correlation function is generally explained in terms of
high-density peak biasing of the galaxy systems (Kaiser 1984), and is
seen in cosmological simulations (\eg, Bahcall \& Cen 1992, Colberg
\etal\  2000 and references therein).  
 
This paper quantifies the angular clustering of pairs of galaxies,
thus exploring the dependence of clustering on system richness in the
regime between single galaxies and groups.
In future work, as an appropriately
large sample becomes available, we will
investigate the clustering of groups of galaxies (see also Lee \& Tucker 2001). 

The Sloan Digital Sky Survey (SDSS; York \etal\ 2000) commissioning
data (Stoughton \etal\ 2001) provide a
photometrically reliable catalog of galaxies over a large field to
\rmag $\approx 22.0$.  We define a uniform sample of compact
galaxy pairs from these data over an area of $278\,\rm deg^2$.  
Scranton \etal\ (2001) show that systematic effects on galaxy
clustering due
to star-galaxy separation, varying seeing, photometric calibration,
reddening, and so on, are small in these data.  The resulting galaxy
clustering analysis is presented by Connolly \etal\ (2001), Tegmark \etal\ (2001),
Szalay \etal\ (2001), and Dodelson \etal\ (2001).   In the present paper, we
extend these results by comparing the angular correlation function of
{\em galaxies} to that of {\em galaxy pairs}.  Our magnitude slice
samples galaxies with a median redshift of roughly 0.22, appreciably
deeper than that of previous group studies. 

 Assuming an appropriate redshift distribution, we invert Limber's
equation to determine correlation lengths for both galaxies and
pairs, and its dependence on the number
density of these systems.  

In \S~\ref{sec:SDSS} we describe the Sloan Digital Sky Survey imaging data, the
properties of the galaxy catalog, and the definition of galaxy
pairs. Estimation of the angular correlation function of galaxies
and of pairs is presented in \S~\ref{sec:wtheta}. In \S~\ref{sec:xi(r)} we invert the angular
function to determine the spatial correlation scales of galaxies and
pairs.  The
dependence of the correlation scale on richness is presented in
\S~\ref{sec:r0-d},
and we give our conclusions in \S~\ref{sec:conclusions}.

\section{The Sloan Digital Sky Survey Imaging Survey}
\label{sec:SDSS}

The Sloan Digital Sky Survey is a photometric and spectroscopic survey
of  1/4 of the sky, above Galactic latitude of $\sim \rm
30^\circ$ (York \etal\ 2000).  The photometric data are
taken with a dedicated 2.5 m altitude-azimuth telescope at Apache
Point, New Mexico, with a $\rm 2.5^\circ$ wide distortion-free field
and an imaging camera consisting of a mosaic of
30 imaging 2048$\times$2048 SITe CCDs with $\rm 0.4''$ pixels (Gunn
\etal\ 1998).  The CCDs are arranged in six columns of five CCDs each, 
using five broad-band filters ($u$, $g$, $r$, $i$ and $z$). The
total integration time per filter is 54.1 seconds.
Each column of CCDs observes a {\em scanline} on the sky roughly
$13'$ wide; the six scanlines of a given observation  make up
a {\em strip}, and two interleaved strips give a {\em stripe}
$2.5^\circ$ wide.  The measured survey depth at which repeat scans
show 
95\% reproducibility is 22.0, 22.2, 22.2, 21.3, 
and 20.5 magnitudes for the 5 filters, respectively.    The SDSS
photometric system is measured in the $\rm AB_{\nu}$ system (Oke \&
Gunn 1983, Fukugita \etal\ 1996).  

Several aspects of the SDSS photometric pipeline (Lupton \etal\ 2001) are
worthy of mention in the context of studies of pairs and groups of galaxies.
We use Petrosian (1976) magnitudes, 
as described by Blanton \etal\ (2001), Yasuda
\etal\ (2001), and Stoughton \etal\ (2001).  
Overlapping images are deblended consistently in five colors, using an
algorithm which makes no assumption regarding profile shape or
symmetry thereof, and which works with an arbitrary number of
overlapping objects.  Visual inspection shows that the morphologies and
photometry of overlapping galaxies are correctly determined for pairs
of objects of similar brightness whose centers are separated by as
little as 3 arcsec. 

Star-galaxy separation is accurate at the 99\% level to at least
$r^*=20.5$\footnote{Because the photometric calibration is not
finalized, we refer to observed photometry with asterisks.  See
Stoughton \etal\ (2001) for a full discussion.}, and better than 90\%
for $r^* \le 22$ (for $\le 1.5''$ seeing); we discuss this
further below.  The data are deep enough that the typical photometric
error in the Petrosian magnitude of an $r^* = 20.8$ galaxy is 0.05
mag, plus 0.03 magnitude error to be added in quadrature due to
uncertainties in the overall photometric calibration. 

\subsection{The Data}
\label{sec:data}

We use imaging data taken during the commissioning period of the SDSS (on
21 and 22 March 1999), which together make up a stripe $2.5^\circ$
wide and $100^\circ$ long  centered on the Celestial Equator (runs 752
and 756); these data are included in the SDSS Early Data Release (Stoughton
\etal\ 2001).   These runs lie within the area $7^h.7 <
{\rm R.A.}\, (2000) < 16^h.8$ and $-1^{\circ}.26 < {\rm Dec.}\, (2000) <
1^{\circ}.26$, although the two strips overlap for only the
central part of this right ascension range. The seeing ranged from $1.2''$ to $2.5''$.

For each of the six scanlines of the two runs, we use the data that is
more than $30''$ from the edge of the CCDs.  We carried out tests of
the robustness of the derived angular correlation function to the
masking of bright stars, and found it to be insensitive to the
masking; in what follows, we do not mask the stars. We carry out
the pair counts on each of the twelve scanlines separately,
normalizing in each one, and then average the results.  Note that this differs
from the approach of Scranton 
\etal\ (2001), who analyze the twelve scanlines together; they also
define a mask which excludes bright stars and regions of particularly
poor seeing.  We will see below that our angular correlation function
is essentially identical to theirs.  The resulting sample has an
effective area of $278.13\,\rm deg^2$.

\begin{figure}
\epsscale{0.7}
\plotone{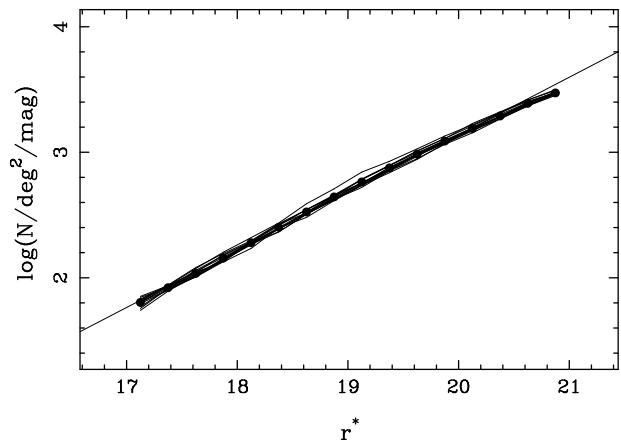}
\caption{\rmag differential galaxy number counts. The data are the number
 of galaxies per magnitude per deg$^2$. The thin solid lines
 represent the counts of 12 individual scanlines. The solid circles
 are the overall number counts in the total $278.13\,\rm deg^2$.
 The bold solid line is a power law fit to the data. The line can be
 represented as ${\log(N\,\rm mag\,\rm deg^2)} = 0.46 r^* -
 6.03$.  The galaxy sample is complete to $r^*=20.5$. 
\label{counts}}
\end{figure}

The turnover in the galaxy number counts as a function of magnitude
(Yasuda \etal\ 2001) provides a good estimate of completeness. The
galaxy number counts in \rmag rise as $d\log N/dm
\approx 0.4$ up to $R<25$ (Infante, Pritchet, \& Quintana 1986; Tyson
1988).  Thus any 
turnover relative to this power law at $r^* < 25$ can be attributed to
incompleteness. Figure~\ref{counts} shows the galaxy counts for our sample,
after correcting for reddening using the extinction maps of Schlegel,
Finkbeiner, \& Davis (1998) (see Yasuda \etal\ 2001 for a thorough
discussion of the SDSS galaxy counts). 
The thin solid lines represent the counts
for each of the 12 scanlines analyzed. The solid dots are the overall
counts for the sample. The bold solid line is the best power law fit
to the data points.  The slope of the counts is $d \log N/dm = 0.46$.
The counts begin to turn over noticeably fainter
than $r^* = 20.5$ (although Scranton \etal\ 2001 show that one can
push star-galaxy separation fainter for galaxy clustering analyses).  We
therefore take our galaxy catalog to be complete to \rmag
$\le20.5$. 

\subsection{The Catalogs}
\label{sec:catalogs}

The next step is to generate catalogs of galaxies and galaxy pairs. 

\begin{figure}
\epsscale{0.7}
\plotone{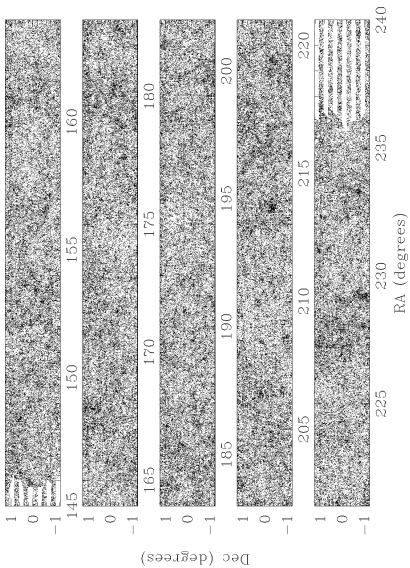}
\caption{Distribution on the sky of galaxies from runs 752 and 756 with
$18 < r^* <20$.  The region shown is centered on the right ascension
range over which the two runs overlap. 
\label{distsky}}
\end{figure}

We wish to work in a fairly narrow magnitude range to minimize
redshift projection effects.  We also want to avoid problems with completeness,
and work in a range where the redshift distribution of galaxies is
well-understood (see below).  With this in mind, we define a sample in
the range $18 \le r^* \le 20$, after correction for reddening as
above; this includes $330,041$ galaxies over an area of $278.13\,\rm
deg^2$.  The distribution on the sky of the galaxies in our sample
is shown in Figure~\ref{distsky}, which shows the region centered on
the right ascension range in which the two strips overlap.  Although
these data are drawn from 12 separate scanlines, the scanline
boundaries are invisible where they overlap.  

The reliability of the catalog depends on photometric uniformity, the
performance of the object deblender, the number of spurious detections
around bright stars and edge effects.  These issues are discussed by
Scranton \etal\ (2001) and Lupton \etal\ (2001). In  brief, overlaps
between adjacent scanlines show photometric zero-point offsets of
less than 0.03 mag; similar conclusions are reached from the agreement
in galaxy counts between the scanlines as shown in
Figure~\ref{counts}.  Visual  examination of pairs and triplets of
galaxies found in the sample (as described below) shows that the
deblender works well in over 95\% of close pairs.  Similarly,
examination of regions around bright stars shows essentially no
spurious objects in the magnitude range we use.  Finally, the overlap
between scanlines (roughly $1'$) means that, for galaxies smaller than
$20''$ (which includes essentially all galaxies in our sample), we can
reject objects which overlap the edges of the CCDs without leaving
any gaps. 

We turn now to describe our criteria for selecting isolated galaxy
pairs.   The galaxy angular correlation function, \wt, quantifies the
ratio of clustered to unclustered systems;  we thus work at
separations where $\omega(\theta) > 1$, to ensure that the excess of
pairs over a random distribution is more than a factor of two (see
Infante, de Mello, \& Menanteau 1996 and Carlberg, Pritchet, \&
Infante 1994 for further discussion).  

\begin{figure}
\epsscale{0.7}
\plotone{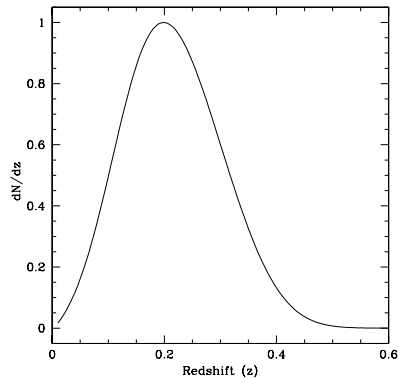}
\caption{Normalized redshift distribution of galaxies in the range
 $18\le r^*\le20$ based on luminosity functions from the
 CNOC2 redshift survey (Lin \etal\ 1999). The median redshift is
 $\langle z\rangle=0.22$, with a FWHM of $0.09$.
\label{dndz}}
\end{figure}

The SDSS galaxy redshift survey (cf.\ Zehavi \etal\ 2001, Stoughton
\etal\ 2001) extends only
to $r^*=17.77$, so we use the luminosity function of the CNOC2
redshift survey by Lin \etal\ (1999) to determine the expected
redshift distribution of the sample.   We carried out 
transformations of the photometric bands, following Fukugita
\etal\ (1996), for each galaxy type (see Dodelson \etal\ 2001 for
further discussion of this point).  The predicted redshift
distribution of galaxies, $dN/dz$, in the range $18\le r^* \le 20$ is
plotted in Figure~\ref{dndz}, and of course is consistent with the
distributions shown by Dodelson \etal\ (2001).   The predicted median
redshift is $\langle z \rangle=0.22$ with a dispersion of  $0.09$.  In
what follows, we explicitly assume that the luminosity function of
galaxies in pairs is the same as that of isolated galaxies. 

We select groups of galaxies as follows; below, we will limit  our
analysis to those groups which have exactly two members. We examine all galaxies in
the catalog, and find all galaxies within an angular separation of
$\theta = 15''$ (in practice, we also require $\theta > 2''$, as the
deblender rarely separates objects reliably at closer
separation). $15''$ corresponds to $37\,h^{-1}$ kpc at the median
redshift of the sample. 
We
then merge all groups  which have members in common.  For each group,
we define the group radius, R$_{G}$, as the angular radius of the smallest
circle containing the centers of the group members (for pairs, R$_{G}$
is half the separation of the two galaxies), and  R$_N$, as the
distance from the group centroid to the next galaxy not in the group.
Our sample of isolated groups is defined by the criterion
R$_{N}$/R$_{G} \ge 3$, which  is based
on those used for local samples (Ka\-ra\-chen\-tsev 1972; Hickson
1982; Prandoni \etal\ 1994), who  show that this decreases the
fraction of chance projections in group-finding.  We note however that our
isolation criterion does not consider galaxies fainter than the sample
limit of $r^* = 20$; in contrast, the low-redshift samples  referenced
above considered all galaxies up to three magnitudes fainter than the
brightest member.  We will show below that this makes little difference
in our analysis.

\begin{figure}
\epsscale{0.7}
\plotone{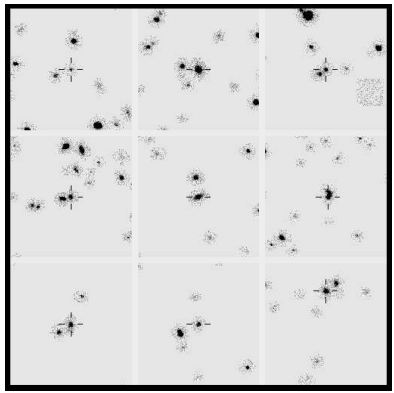}
\caption{Examples of galaxy pairs in \rmag. The cross-hair marks one 
of the members. Each panel is $1.34\times1.34~\rm arcmin^2$.  These
images include foreground stars as well, and are recreated from the
SDSS ``Atlas Images'', postage stamps extracted around each detected
object.  Thus there is no background sky noise beyond the boundary of
each object. 
\label{pairs-examples}}
\end{figure}

This sample contains 15,492 close
pairs, but only 1175 groups with three or more members, too small a
sample to measure a robust correlation function.  We therefore carry
out the clustering analysis only on the pairs.  We 
inspected the images of a large number of these pairs. 
Fewer than 3\% are spurious detections, \eg
improperly deblended bright extended objects or bright star spikes,
meteor trails, \etc. Fig. \ref{pairs-examples} shows some examples of
galaxy pairs from our sample.

In Table 1 we present the main characteristics of the galaxy and pair
catalogs.  We explicitly assume that the luminosity function,
and thus the redshift distribution of galaxies in pairs is the same as that of field
galaxies.  We list the mean redshift and effective width of the
$dN/dz$ distribution.  The effective volume, weighted by $dN/dz$,
is listed for the samples (equation~\ref{eq:v_eff}).  The correlation
statistics are described below.

\section{The Angular Correlation Function of Galaxies and Pairs}
\label{sec:wtheta}

We determine the angular correlation function of galaxies and of pairs
of galaxies using a catalog of randomly distributed points over the
survey area, and using the estimator suggested by Landy \& Szalay (1993):

\begin{equation}
\omega (\theta )={{N_{dd} - 2~N_{dr} + N_{rr}}\over{N_{rr}}}.
\end{equation}

\noindent
Here $N_{dd}$ is the number of galaxy pairs in a given range of
separations summed over all SDSS fields, $N_{dr}$ is the number of
galaxy-random pairs, and  $N_{rr}$ is the number
of random pairs.  Pair counts were done in each
scanline separately. The random pair counts were obtained from
100 random catalogs, with galaxies randomly distributed
within the net survey area. Random counts were normalized to each
scanline separately. 
We also calculated the correlation function using the method of
Infante (1994); the results are indistinguishable. 

\begin{figure}
\epsscale{0.7}
\plotone{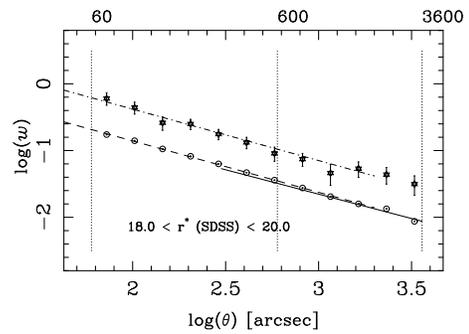}
\caption{Angular correlation function of galaxies and galaxy pairs
for $18\le r^* \le 20$.  The x-axis is separation in arcsec, measured
on a logarithmic scale on the lower axis, and a linear scale on the
upper axis.  The open points are the measured galaxy
correlation function, and the solid points are the measured pair
correlation function.  The thin solid line is the fit to the Connolly \etal\
(2001) correlation function with $\gamma=1.73$, scaled to our 
magnitude range.  The dashed line is the
best fit model for galaxies, with $r_0=4.2~h^{-1}$ Mpc.  The dot-dashed
line is the best fit model for pairs with $r_0=7.8~h^{-1}$ Mpc.  The
error bars are determined by bootstrap resampling, and are smaller
than the symbol size for the galaxies.  
\label{w.18-20}}
\end{figure}

\begin{figure}
\epsscale{0.7}
\plotone{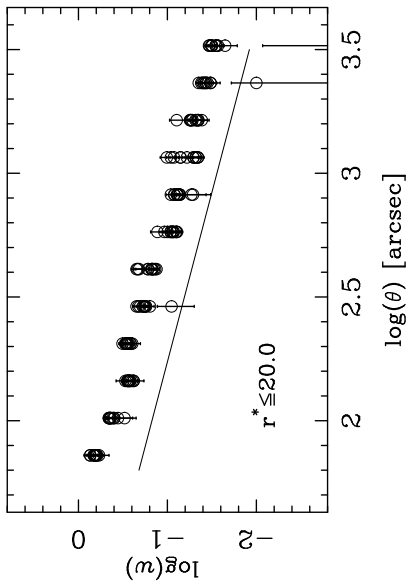}
\caption{Angular correlation function of galaxy pairs
for each scanline separately.  The solid line is the fit to the 
SDSS galaxy correlation function from Connolly \etal\ (2001), scaled
to our magnitude range. 
\label{plcc}}
\end{figure}

The resulting angular correlation functions of galaxies and galaxy
pairs are shown in Figure~\ref{w.18-20} and Table 1.  The error bars
result from 100 bootstrap re-sampling variations; these conservative
errors (see the discussion in Fisher \etal\ 1994) are somewhat
larger than the scatter from one scanline to another, presented for
the galaxy pair sample in Figure~\ref{plcc}.  The galaxy angular
correlation function follows a power law, $\omega =
A_{\omega}\theta^{(1-\gamma)}$, with $\gamma = 1.77 \pm 0.04$, and an
amplitude essentially identical with that given by Connolly \etal\
(2001), when their results for the $18 < r^* < 19$ and $19 < r^* < 20$
slices are combined.   In agreement with the latter paper, we find no
breaks in the correlation function on scales smaller than one degree. 

The galaxy pair correlation function shows the same power law behavior
as the galaxies, but with a significantly larger amplitude.  

The error bars determined by the bootstrap resampling method in the
galaxy correlations are smaller than the symbols.  Scranton \etal\
(2001) discuss the angular correlation and the related uncertainties
in greater detail; our results are consistent with theirs.  We have
also carried out a jack-knife estimation of the errors, and find
consistent results with those presented here. 

  We fit each of the correlation functions to a power law; the results
are given in Table 1.  These fits use the bootstrap errors, but unlike
Connolly \etal\ (2001) do not take the covariance into account; we may
therefore be underestimating our error bars on $r_0$ somewhat.  

  Our galaxy pair sample is defined without redshifts, over a fairly
narrow range of apparent magnitude.  This could give rise to two types
of systematic error in the estimated clustering strength.  First,
distant clusters of galaxies might have only two members
bright enough to enter the galaxy sample, and therefore could
masquerade as a galaxy pair; as clusters exhibit appreciably stronger
clustering than do groups (e.g., Bahcall 1988), this could bias the
results.  However, at $z\sim0.22$, the density of galaxy pairs is much
higher than the density of clusters.  Thus, any contamination from
background clusters should be small. To check this, we searched for
galaxies as faint as $r^*=21.5$ around each pair, and found no
significant density enhancements.  Thus there is no evidence that many
of our pairs are the ``tips of the iceberg'' of distant clusters.

  Second, inclusion of pairs in projection will tend to systematically
decrease the clustering amplitude.  We have carried out simulations of
the effects of random projection on the clustering of our pair
catalog.  We find that of order $f = 30\%$ of our pairs are likely to
be chance projections along the line of sight.  The angular
correlation function is depressed by a factor $(1-f)^2$ by projection;
this suggests that the correlation amplitude of groups listed in Table
1 should be multiplied by a factor of order two.

\section{The Three-Dimensional Correlation Function of Galaxies and Pairs}
\label{sec:xi(r)}

We relate our \wt\ measurements to the three-dimensional
correlation functions $\xi(r)$ via Limber's Equation (Limber 1953). The spatial correlation
function is assumed to be a power law weighted by the standard
phenomenological evolutionary
factor, $\xi(r,z) =
\left({r}\over{r_0}\right)^{-\gamma}(1+z)^{-(3+\epsilon)}$, 
where $r$ is the proper distance, $r_0$ is the proper correlation
length, and $\epsilon$ is the clustering evolution index.  
$\epsilon=\gamma-3$ corresponds to  clustering fixed in comoving
coordinates, while
$\epsilon=0$ represents stable clustering in physical coordinates
(Phillips \etal\ 1978).

Parameterizing the angular clustering as $\omega(\theta) = A_\omega
\theta^{(1-\gamma)}$, $r_0$ is given by

\begin{equation} 
r_0^{\gamma}=A_\omega^{-1}~C~{{\int_{0}^{\infty} g(z) (dN/dz)^2 dz } \over 
{\lbrack \int_{0}^{\infty} (dN/dz) dz\rbrack^2 }}
\end{equation}
(\eg Efstathiou \etal\ 1991; see also Peebles 1980 and Phillips 
\etal\ 1978). $C$ is a constant involving purely numerical
factors, 
\begin{equation} 
C=\pi^{1/2}~{{\Gamma [(\gamma -1)/2]}\over{\Gamma (\gamma /2)}}
\end{equation}
and the function $g(z)$ depends only on $\epsilon$, $\gamma$,
and cosmology: 
\begin{equation} 
g(z)=\left({dz}\over{dx}\right)x^{1-\gamma}F(x)(1+z)^{-(3+\epsilon-\gamma)}
\end{equation}
where $x(z)$ is the coordinate distance and $F(x)$ is
\begin{equation} 
F(x)=[1-(H_0a_0x/c)^2(\Omega_0-1)]^{1/2}.
\end{equation}

We use a cosmology with $q_0=0.1$ and $\Lambda=0$, and an evolution
index of $\epsilon=0$ as suggested by observations (Carlberg \etal\ 2000).

The $H_0$ dependence of $g(z)$ is canceled by that in the
measured value of $r_0$. The strong dependence of $A_\omega$ on
$dN/dz$ is clear in this equation. Note in particular that the
above equations {\it are not affected by galaxy evolution, 
except in the calculation of the redshift distribution $dN/dz$}
(Peebles 1980, eqs. [56.7] and [56.13]).  However, Limber's
equation, as  presented here, does assume that galaxy clustering
is independent of luminosity; the high-redshift objects in our narrow
magnitude slice are of higher luminosity than the low-redshift
objects.  Clustering is in fact weakly dependent on luminosity (e.g.,
Norberg \etal\ 2001 and references therein), but the quantification of
this effect on our results is beyond the scope of this paper. 

Because our sample covers a narrow magnitude range, and the
galaxies in pairs have essentially the same mean magnitude as the
entire galaxy sample (Table 1), we expect the $dN/dz$ of pairs to be
identical to that of galaxies (Figure~\ref{dndz}).  This
allows us to estimate the  correlation
lengths, $r_0$, for the clustering of galaxies and pairs. We find that
pairs  have a significantly larger correlation length than do
galaxies; $r_0=4.2\pm0.4$ $h^{-1}~\rm Mpc$ for galaxies and
$7.8\pm0.7 h^{-1}~\rm Mpc$ for pairs.  These values are to be compared
with $\sim 5$ $h^{-1}~\rm Mpc$ for galaxies (Davis \& Peebles 1983;
Loveday \etal\ 1996, Infante \& Pritchet 1995), $\sim 7-8$ $h^{-1}~\rm
Mpc$ for groups  (Carlberg \etal\ 2001, Girardi \etal\ 2000, Merch\'an
\etal\ 2000) and $\sim 15-20$ $h^{-1}~\rm Mpc$ for rich clusters (Bahcall 
\& West 1992).  The errors in our derived $r_0$ include both
statistical uncertainty, as well as an uncertainty added
in quadrature due to our imperfect knowledge of the redshift
distribution (cf., the discussion by Dodelson \etal\ 2001). 

In the following section, we will relate the clustering length to the
mean inter-system separation of galaxies and groups.  We determine the
latter from the effective volume of the sample, based on the model
redshift distribution: 
\begin{equation} 
V = {{\int_{z_{min}}^{z_{max}}  (dN/dz)  V(z) dz } \over 
{\int_{z_{min}}^{z_{max}} (dN/dz) dz }}
\label{eq:v_eff}
\end{equation}
which then yields the number density, $n = {N_{systems} \over {V}}$, and mean separation, 
$d = \left({1} \over {n}\right)^{1/3}$. 
We find the mean separation of galaxies and galaxy pairs is 3.7 and
10.2 $h^{-1}$ Mpc, respectively.  Note that random superpositions will
artificially depress the pairs value of $d$ by a factor $(1-f)^{1/3}
\approx 0.9$; thus the true value of $d$ for pairs is probably closer
to $11\,h^{-1}$ Mpc. 

\section{The $r_0~-~d$ Relation}
\label{sec:r0-d}

\begin{figure}
\epsscale{0.7}
\plotone{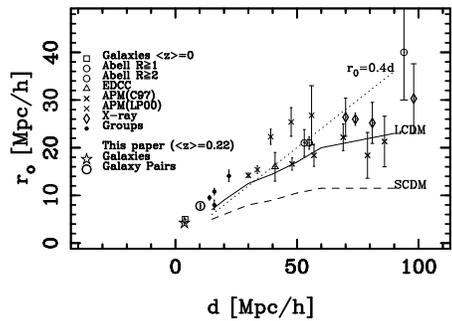}
\caption{Correlation length $r_0$ versus mean separation $d\, (=n^{-1/3})$
for galaxies and pairs, as well as groups and clusters of galaxies
(see \S~\ref{sec:r0-d}, which describes the different samples
plotted). The predictions from simulations of two 
cosmological models, LCDM and SCDM, are shown for 
comparison, as well as the original approximate relation of 
$r_0~\approx~0.4~d$ for rich systems (Bahcall 1988).
\label{plr0-d}}
\end{figure}

The correlation lengths of galaxies and galaxy pairs
are compared with the global richness-dependent correlations of galaxy
systems  (Bahcall \& Soneira 1983, Szalay \& Schramm  1985, Bahcall
1988) in Fig.~\ref{plr0-d}.  Here the correlation scales $r_0$ of
various systems are plotted as a function of the mean separation of
the objects, $d$; since richer systems are rarer, the quantity $d$
scales with the richness or mass of the systems.  The data points
include groups and clusters of galaxies of different richnesses and
different samples: rich Abell clusters (Bahcall \& Soneira 1983,
Peacock \& West 1992, Postman \etal\ 1992, Lee \& Park 2000; 
note that Richness 0
clusters cannot be included in any statistical analyses since they are
no complete samples of these objects); APM clusters (Croft \etal\
1997 (C97);  Lee \& Park 2000 (LP00); the latter find considerably larger
$r_0$ in their re-analysis of  the APM clusters); EDCC clusters
(Nichol \etal\ 1992):  X-ray selected clusters
(REFLEX survey, B\"ohringer \etal\ 2001; XBACS survey, Abadi \etal\
1998, Lee \& Park 2000); and groups of galaxies (Merch\'an \etal\ 2000; Girardi \etal\
2000). The well-known dependence of $r_0$ on $d$ (Bahcall 1988,
Bahcall \& West 1992) is clearly observed. The new result presented
for pairs  of galaxies, with $d = 10.2~h^{-1}$ Mpc and $r_0 = 7.8 \pm
0.7~h^{-1}$ Mpc, fits well within this universal clustering trend; it
is significantly larger than the galaxy correlation scale, and somewhat
smaller than the correlation
scales observed for small groups of galaxies. 

Also presented in Fig. \ref{plr0-d} is the expected $r_0~ -~ d$
relation for two cosmological models: LCDM ($\Omega_m=0.3$,
$\Omega_L=0.7$, $h=0.7$) and Standard CDM (SCDM; $\Omega_m=1$,
$h=0.5$) (both models are normalized to the present-day cluster
abundance). These relations are obtained from large scale, high
resolution cosmological simulations (Governato \etal\ 2000; Colberg
\etal\ 2000; Bahcall \etal\ 2001).  Figure~\ref{plr0-d}
highlights the fact that the SCDM model is highly inconsistent with
the clustering data -- not only for rich clusters, as was previously
known (e.g., Bahcall \& Cen 1992, Croft \etal\ 1997, Governato \etal\
2000, Colberg \etal\ 2000) -- but also for small groups and galaxy
pairs; the correlation strength observed for all these systems is
systematically stronger than predicted by SCDM models. The LCDM model
provides a considerably better match to the data; the lower the value
of $\Omega_m$, the stronger the group correlations predicted.  This `canonical'
model is consistent with most other observations of large scale
structure, clusters of galaxies, supernovae Ia, and the cosmic
microwave fluctuations (\eg, Bahcall \etal\ 1999, Wang \etal\ 2000).
In fact, the data may even suggest a somewhat
lower $\Omega_m$ than 0.3, as some of the best data points exhibit
stronger correlations than expected for $\Omega_m = 0.3$.

Note that the lower $r_0$ values for the two richest (but small)
sub-samples of APM clusters are due to their exceptionally steep
correlation function slope: $\gamma\sim -3$ instead of $\sim -2$; all
other sample have a slope of $\sim -2$. This steepness reduces $r_0$,
bringing it lower than the other relevant data.

\section{Conclusions}
\label{sec:conclusions}

We use a sample of 330,041 galaxies within a stripe of area 278 deg$^2$ centered on 
the Celestial Equator, with magnitudes $18\le r^* \le 20$, obtained
from SDSS commissioning imaging data.  We use these data to select
isolated pairs of galaxies. We determine the angular correlation
function of the galaxies and of the galaxy pairs.
We find the following  results:

\begin{itemize}

\item   Pairs of galaxies are more strongly clustered than are single
galaxies.   
The angular correlation amplitude of galaxy pairs is
 $2.9\pm0.4$ times larger than that of galaxies.

\item The power-law slopes of the two correlation functions are the
same, corresponding to 
$\gamma = 1.77\pm 0.04$. 

\item We measure \wt\ up to 1 deg scales, corresponding to $\sim9~
h^{-1}~\rm Mpc$ at the mean redshift of 0.22. No breaks  are detected in either
correlation function.

\item Assuming a redshift distribution from the CNOC2 survey
luminosity function, we
invert the angular correlations and determine a three-dimensional
correlation length, $r_0$, for each sample. We find $r_0 = 4.2\pm0.4$
$h^{-1}~\rm Mpc$ for galaxies and $7.8\pm0.7 h^{-1}~\rm Mpc$ for
pairs; the latter may be biased downward somewhat by projection
effects. 

\item The mean separation between systems is
$d = 3.7$ and $10.2~h^{-1}$ Mpc for galaxies and pairs
respectively. The correlation lengths fit the global $r_0~-~d$ relation
observed for galaxy systems well (Bahcall \& West 1994).
\end{itemize}

  This work suggests a number of follow-up studies.  We have defined
pairs of galaxies from their positions on the sky, and we need to
quantify what fraction of these objects are pairs at the same
redshift.  We are in 
the process of carrying out a redshift survey of a subset of the pairs
sample, which will also be useful in tying down the $dN/dz$ relation.
We can also use photometric redshifts from the multi-band photometry
of the SDSS to define a cleaner sample of galaxy pairs.   Finally, we have used just under 300
deg$^2$ of SDSS data in this study; the survey has now imaged
over five times this area.  Thus we are in the process of defining
samples of 
richer (and thus rarer) systems from this larger area, and measuring
their correlations, to investigate the $r_0-d$ relation between for
richer groups.

\acknowledgments
The Sloan Digital Sky Survey (SDSS) is a joint project of The
University of Chicago, Fermilab, the Institute for Advanced Study, the
Japan Participation Group, The Johns Hopkins University, the
Max-Planck-Institute for Astronomy (MPIA), the Max-Planck-Institute
for Astrophysics (MPA), New Mexico State University, Princeton
University, the United States Naval Observatory, and the University of
Washington. Apache Point Observatory, site of the SDSS telescopes, is
operated by the Astrophysical Research Consortium (ARC).

Funding for the project has been provided by the Alfred P. Sloan
Foundation, the SDSS member institutions, the National Aeronautics and
Space Administration, the National Science Foundation, the
U.S. Department of Energy, the Japanese Monbukagakusho, and the Max
Planck Society. The SDSS Web site is http://www.sdss.org/.   We thank
Scott Dodelson and Douglas Tucker for useful comments on the text.  LI is
grateful to Fundaci\'on Andes for financial support, to Princeton
University for its hospitality, FONDECYT and to the Guggenheim Foundation. MAS
acknowledges the support of NSF grants AST-9616901 and AST-0071091,
and MSV acknowledges the support of NSF grant AST-0071201.


\begin{deluxetable}{lcc}
\tablewidth{0pc}
\tablecaption{Properties of Galaxies and Pairs ($18 \le r^* \le 20$)} 
\tablehead{
\colhead{Property} & \colhead{Galaxies} & \colhead{Pairs} }
\startdata
Mean redshift $\langle z\rangle$       & $0.22\pm0.1$ &$0.22\pm0.1$      \\
Effective Volume [$h^{-3}~\rm Mpc^3$]  & $17.8\times10^6$ & $17.8\times10^6$ \\  
$\langle r^*\rangle$ [SDSS band]      & 19.34            & 19.30          \\
Number              & 330,041    & 15,492   \\
$A_\omega$, ($\gamma$)   & $4.94\pm0.02$ (1.77)& $13.54\pm0.07$ (1.76)\\
Correlation length $r_0~[h^{-1}~\rm Mpc]$       & $4.2\pm0.4$         & $7.8\pm0.7$    \\
Mean spacing $d\ [h^{-1}~\rm Mpc]$        & 3.7              & $10.2$   \\
\tablecomments{$H_0 = 100h\rm \, km\,s^{-1}\,Mpc^{-1}$, 
$q_0=0.1$, $\Lambda=0$, $\epsilon=0$;\\
Area covered: $278.13~\rm deg^2$}
\enddata
\end{deluxetable}


\end{document}